\documentstyle[12pt]{article}
\newcommand{\eq}{\begin{equation}}
\newcommand{\en}{\end{equation}}
\newcommand{\EQ}{\begin{equation}}
\newcommand{\EN}{\end{equation}}
\newcommand{\bea}{\begin{eqnarray}}
\newcommand{\ena}{\end{eqnarray}}

\renewcommand{\a}{\alpha}

\begin{document}

\begin{titlepage}

\large
\baselineskip=16pt

\hfill {\small \bf DSF-T-43/93}

\hfill {\small \bf INFN-NA-IV-43/93}

{\Large \bf
\begin{center}
On the Zero-Slope Limit \\
of the Compactified Closed Bosonic String\footnote{This work is carried
out in the
framework of the E.C. Research Programme ``Gauge Theories, Applied
Supersymmetry
and Quantum Gravity'', under contract SCI-CT92-0789.}
\end{center}}

\begin{center}
R. Marotta$^{a,b}$ and F. Pezzella$^{b}$ \

\vspace{0.2cm}

{\normalsize $^{a}$ Dipartimento di Scienze Fisiche,  Universit\`a di Napoli\\
Mostra d'Oltremare, Pad. 19, I-80125  Napoli, Italy}\\
\vspace{0.1cm}
{\normalsize
$^{b}$ I.N.F.N. - Sezione di Napoli \\
Mostra d'Oltremare, Pad. 20, I-80125 Napoli, Italy} \\
\vspace{0.1cm}
\end{center}

\vspace{0.5cm}

{\small
\centerline{\bf Abstract}

\begin{quote}

In the framework of the compactified closed bosonic
string theory with the extra spatial coordinates being circular
with radius $R$, we perform both the zero-slope limit
and the $R \rightarrow 0$ limit of the tree scattering amplitude
of four massless scalar particles. We
explicitly show that this double limit leads to amplitudes involving
scalars which interact through the exchange of a scalar, spin $1$ and
spin $2$ particle. In particular, this latter case reproduces the same
result obtained in linearized quantum gravity.
\end{quote} }

\end{titlepage}
\setcounter{footnote}{0}
\newpage
${\bf 1.}$
String theories have to reproduce, at the ``low-energy'' limit in which the
slope $\a ' $ goes to
zero, ordinary field theories
formulated in a number of
space-time dimensions, which coincides with the dimensionality $D$
of the space-time in which the string is embedded ($D=26$ for the
bosonic string). In order to reduce this
dimension to $D=4$ a compactification scheme must be adopted. Our
work just goes in the direction to show explicitly how, with a suitable
compactification procedure, string amplitudes lead
to the ones of four-dimensional field theories in the above mentioned limit.

We will consider here the so-called ``toroidal compactification'' \cite
{GSW} \cite{LT} \cite{LSW} of the
bosonic closed string,
in which twenty-two spatial coordinates are compactified into
circles with radius $R$ and the compactified space becomes
a lattice which is required to be Lorentzian, self-dual and even in order
to have consistency with the properties of the bosonic string theory.

Constructing such a lattice yields the introduction of Lie algebra lattices
where the massless states which arise from the toroidal compactification
lie on the root lattice and belong to the adjoint representation
of the gauge group relative to the algebra.

After having endowed the theory with the above compactification procedure,
the
limit $\a ' \rightarrow 0$ makes all the massive modes uncouple
giving rise to a description in terms of only massless states.

In particular we consider tree scattering amplitudes
of massless scalar particles in the compactified bosonic
closed string theory. We perform the double limit $\a ' \rightarrow 0$
and $R \rightarrow 0$, keeping the ratio $a=R/\sqrt{\alpha'}$ fixed;
in so doing,
we obtain amplitudes corresponding
to the exchange of a scalar, spin 1 and spin 2 particle. In particular
this latter amplitude is coincident
with the one obtained in the
framework of linearized quantum gravity.
An analogous computation was performed in the context of the
generalized dual Virasoro model \cite{Y}.
\vspace{0.5cm}

${\bf 2.}$ The toroidal compactification consists in associating the
$d$ internal extra-coordinates of the string to $d$ circles having radius
$R_{i}$ with $i=1, \dots , d$; this can be done by
identifying the points of the internal space as follows:
\vspace{0.2cm}
\[
X^{I} \equiv X^{I} + 2 \pi L^{I}
\]
where $I =  1, \dots, d $ and
\[
L^{I}= \sqrt \frac{1}{2} \sum_{i=1}^{d} n_{i} R_{i} e_{i}^{I}
\]
being $ n_{i} \in Z $ a so-called ``winding number''. The vectors
$\hat{e}_{i} \equiv (e_{i}^{1},
\dots, e_{i}^{d})$ are linearly independent and
normalized as follows:
\[
\hat{e}_{i} \cdot \hat{e}_{i} = 2 .
\]

The quantities $ L^{I} $'s can  be thought as the components
of a vector
defined on a $d$-dimensional lattice $\Lambda^{d}$ which admits as a basis
the set of vectors $\left\{
\sqrt{\frac{1}{2}} R_{i} \hat{e}_{i}
\right\}$ with $i=1, \dots, d$.
It follows that the torus on which we compactify is the quotient
space:
\[
                  T^{d} = \frac{ {\mbox {\bf R}}^{d}}{2 \pi \Lambda^{d}}.
\]
One gets the following mode expansion for the compactified
string field $X^{I}$
\cite{LT} \cite{LSW}:
\[
X^{I}(z, \bar{z}) = X_{L}^{I}(z) + X_{R}^{I}(\bar{z})
\]
with
\bea
X_{L}^{I}(z) & = & x^{I}_{L} - i \frac{\a'}{2} p^{I}_{L} \,\,\, \mbox{log} z
+ i \sqrt{\frac{\a'}{2}} \sum_{n \neq 0} \frac{1}{n} \a^{I}_{n}
z^{-n}     \\
X_{R}^{I}(\bar{z}) & = & x^{I}_{R} - i \frac{\a'}{2} p^{I}_{R} \,\,\,
\mbox{log} \bar{z} + i
\sqrt{\frac{\a'}{2}} \sum_{n \neq 0} \frac{1}{n} \bar{\a}^{I}_{n}
\bar{z}^{-n}           \label{xrl}
\ena
where
\[
x_{L}^{I}= \frac{1}{2} x^{I} + \frac{\a'}{2} Q^{I}
\]
\[
x_{R}^{I}= \frac{1}{2} x^{I} - \frac{\a'}{2} Q^{I}
\]
\vspace{0.2cm}
being $Q^{I}$ the operator canonically conjugate to $L^{I}$, here
introduced in order to define completely
independent left and right sectors~\cite{LSW}; furthermore,
\vspace{0.2cm}
\[
p^{I}_{R} = p^{I} - \frac{L^{I}}{\a'} \,\,\,;\,\,\,
p^{I}_{L} = p^{I} + \frac{L^{I}}{\a'}.
\]
\vspace{0.2cm}
The following commutation relations hold:
\[
\left[ x^{I}_{L}, p^{J}_{L} \right] = \left[ x^{I}_{R}, p^{J}_{R}
\right] = i \delta^{IJ}
\]
with all the other commutators vanishing.
The compactification of the internal spatial coordinates implies that
also the
momenta $p^{I}$'s, which represent the translation operators of those
coordinates, lie on a $d$-dimensional lattice that is the dual of the
lattice $\Lambda^{d}$ and it is denoted by $( \Lambda^{d})^{*} $, i.e.:
\eq
p^{I}=\sqrt{2} \sum_{i=1}^{d} \frac{m_{i}}{R_{i}} e_{i}^{*I}
\en
being the vector $\hat{e}_{i}^{*}$ the dual of $\hat{e}_{i}$. A basis
on such a lattice is given by the vectors $
\left\{
\frac{\sqrt{2}}{R_{i}} \hat{e}_{i} \right\}$.
In this compactification scheme, in which we consider
$R_{i} = R$ $\forall i=1,\dots,d$, the constraint conditions of
the bosonic closed string become:
\vspace{0.2cm}
\eq
L_{0} - 1 = 0   \Longleftrightarrow \frac{\a'}{4} m^{2} = \frac{\a'}{4}
p^{2}_{R} + N - 1
                                 \label{const1}
\en
\eq
\bar{L}_{0}-1=0 \Longleftrightarrow \frac{\a'}{4} m^{2} = \frac{\a'}{4}
p^{2}_{L} + \bar{N} - 1
                                           \label{const2}
\en
The conditions (\ref{const1}) and (\ref{const2}) can be rewritten as
follows:
\bea
\frac{\a'}{2} m^{2} &  = & \frac{\a'}{4} \left( p^{2}_{R} + p^{2}_{L}
\right)
+ N + \bar{N} - 2  \label{1}\\
N - \bar{N} & = & \frac{\a'}{4} \left( p^{2}_{L} - p^{2}_{R} \right) .
\label{2}
\ena
{}From here it is possible to observe that the bivector
\[
\hat{P} \equiv \left( \sqrt \frac{\a'}{2} p_{R}, \sqrt \frac{\a'}{2} p_{L}
\right)
\]
lies on an even lattice $\Gamma_{d,d}$, after having chosen the metric
of
the lattice to be of the form $((+1)^{d},(-1)^{d})$ (Lorentzian lattice);
furthermore the
modular invariance imposes that such a lattice must be self-dual too.

Another condition that must be imposed on the lattice comes out from
the following considerations.

By analyzing the lattice $\Gamma_{d,d}$ it turns out that the
right and the left components of the bivector $\hat{P}$  can be written as:
\eq
\sqrt \frac{\a'}{2} p_{R}
 = \sum_{i=1}^{d} m_{i} \left( \frac{\sqrt{\a'}}{R} \right)
\hat{e}_{i}^{*} - \frac{1}{2} \sum_{i=1}^{d} n_{i} \left(
\frac{R}{\sqrt{\a'}} \right) \hat{e}_{i},   \label{pr}
\en
\eq
\sqrt \frac{\a'}{2} p_{L}
 = \sum_{i=1}^{d} m_{i} \left( \frac{\sqrt{\a'}}{R} \right)
\hat{e}_{i}^{*} + \frac{1}{2} \sum_{i=1}^{d} n_{i} \left(
\frac{R}{\sqrt{\a'}} \right) \hat{e}_{i} .  \label{pl}
\en

The expressions (\ref{pr}) and (\ref{pl})
can be generalized by adding a constant
background anti-symmetric tensor field $B_{ij}$ to the usual action of the
bosonic string: this operation is necessary to get
more general and larger gauge groups \cite{LT} \cite{LSW}.
Taking into account this generalization we
can rewrite the components of the bivector $\hat{P}$ as follows:
\[
\sqrt \frac{\a'}{2} p_{R}^{I}
 = \sum_{i=1}^{d} m_{i} \left( \frac{\sqrt{\a'}}{R} \right)
{e_{i}}^{*I} - \frac{1}{2} \sum_{i=1}^{d} n_{i} \left(
\frac{R}{\sqrt{\a'}} \right) {e}_{i}^{I} - \sum_{ij=1}^{d} B_{ij}
n_{j} \left( \frac{\sqrt{\a'}}{R} \right) e_{i}^{*I}
\]
\[
\sqrt \frac{\a'}{2} p_{L}^{I}
 = \sum_{i=1}^{d} m_{i} \left( \frac{\sqrt{\a'}}{R} \right)
\hat{e}_{i}^{*I} + \frac{1}{2} \sum_{i=1}^{d} n_{i} \left(
\frac{R}{\sqrt{\a'}} \right) \hat{e}_{i}^{I} -\sum_{ij=1}^{d} B_{ij}
n_{j} \left( \frac{\sqrt{\a'}}{R} \right) e_{i}^{*I}
\]
The vector basis in $\Gamma_{d,d}$ are evidently $\frac{\sqrt{\a'}}{R}
\hat{e}_{i}^{*}$ and the dual ones are $\frac{R}{\sqrt{\a'}} \hat{e}_{i}$.

In the double limit $\a' \rightarrow
0$, $R \rightarrow 0$,  $p_{L}$ and $p_{R}$ will be well-defined
quantities only if
the ratio $a=\frac{R}{\sqrt{\a'}}$ is kept fixed \cite{GS}; in particular we
choose $a=1$ \cite{GSW}. This choice leads to a rational lattice.

In conclusion, the lattice on which we compactify must be Lorentzian,
self-dual, even and rational.

It is known that a large class of such lattices can be constructed in
${\mbox {\bf R}}^{d,d}$ considering the set of all vectors of the form
$(v_{1}, v_{2})$ so that $v_{1}$ and $v_{2}$ belong to the same
conjugacy class of a semi-simple Lie algebra of rank $d$ \cite{LSW}.

By evaluating the double limit $R = \sqrt{\a'} \rightarrow 0$
on the equations (\ref{1}) and (\ref{2})
one has that
the only particles with finite masses which survive
are massless particles for which the norm of the components
of $\hat{P}$ is null or equal to $2$. In particular, these latter
are by
definition lattice roots. The roots include in any case
those
of the Lie algebra used in constructing the lattice, but in some cases
there may be additional norm 2 vectors in the other coniugacy classes.

We are going to compactify on a lattice where all the norm 2 vectors
belong to the root lattice of a simply laced Lie algebra. On this kind of
lattices
the scalar product between the vectors, which survive after performing
the double limit,
takes integer values. This will greatly simplify our computation.

A possible lattice satisfying the above requirements is $\Gamma_{d,d} =
\Gamma_{d} \otimes \Gamma_{d}$ with $\Gamma_{d}= E_{8} \otimes E_{8}
\otimes E_{6}$; however there exist a large class of
Lorentzian, self-dual, even and rational lattices that well suite to
our problem.

\vspace{0.5cm}

{\bf 3.} We are going to consider scattering amplitudes involving
scalar particles. These come from the levels reported in
table 1, together with the corresponding norms of
$p_{R}$ and $p_{L}$.

In order to compute those scattering amplitudes, we are going to consider a
compactified version
of the vertex operators $V_{\psi}(z, \bar{z})$ providing
the amplitude for the emission of a state $\psi$ in the string spectrum.

The compactified vertices can be obtained from the ordinary ones through
simple correspondences; for example, the compactified vertex relative to
the scalar particles belonging to the level $N=\bar{N}=0$ is the naive
version of the compactified tachyon vertex:
\[
V_{0} \left( k,k_{R},k_{L}; z, \bar{z} \right) =
: e^{i k X(z, \bar{z})} e^{i k_{R} X_{R} (z)} e^{ i k_{L} X_{L} (\bar{z})} :
\]
where $X^{\mu} (z, \bar{z})$ is the usual string field, with
$\mu=1,\cdots, 26-d$; $X^{I}_{R}$ and $X^{I}_{L}$, with $I=1, \dots, d$
are the fields defined in (\ref{xrl}). The state corresponding to $V_{0}$
is defined, as usual, through the following limit:
\[
\lim_{z, \bar{z} \rightarrow 0}  V_{0} (k, k_{R}, k_{L};z, \bar{z} )
|\mbox{vacuum}>  .
\]

Conformal invariance requires $V_{0}$ to be a conformal field with
dimensions $\Delta=\bar{\Delta}=1$; since the following operator product
expansion holds between $V_{0}$ and the stress energy tensor:
\[
T(z) V_{0} (w,\bar{w}) = \left[ \frac{1}{z-w} \partial_{w} +
\frac{\left( k^{2} + k^{2}_{R} \right) \frac{\a'}{4}}{(z-w)^{2}} \right]
V_{0}(w, \bar{w})
\]
it follows that
\[
k^{2} + k^{2}_{R} = \frac{4}{\a'}
\]
with a similar relation holding for the anti-holomorphic sector.
This shows again that for massless scalar particles one has $k^{2}=0$ and
$k^{2}_{R}=k^{2}_{L}=\frac{4}{\a'}$.

These considerations suggest to introduce the following general
vertex, which is nothing but a linear combination
of the vertices relative to the massless scalar
particles introduced in table 1:
\eq
V_{s} = : e^{ikX(z, \bar{z})} \left[ e^{ i k_{R} X_{R}(z)} + \xi \cdot
\partial_{z} X_{R}(z) \right] \left[ e^{i k_{L} X_{L}(\bar{z})} + \bar{\xi}
\cdot
\partial_{\bar{z}} X_{L}(\bar{z}) \right] :  \label{Vs}
\en
where $\xi$ and $\bar{\xi}$ are polarization vectors defined in the
compactified space.

By using the version (\ref{Vs}) of the vertex operators
it is straightforward to compute the tree scattering amplitude of four
scalar particles in $D$ dimensions, by using the operatorial formalism of
the $N$-String Vertex~\cite{PDV} [see fig. 1]:
\eq
A= N_{0} \int dz d\bar{z} \,\, z^{ \frac{\a'}{2} k_{3}k_{4} + n_{34} }
(1-z)^{ \frac{\a'}{2} k_{2}k_{3} + n_{23}} \bar{z}^{ \frac{\a'}{2} k_{3}k_{4} +
\bar{n}_{34} } (1-\bar{z})^{ \frac{\a'}{2} k_{2}k_{3} + \bar{n}_{23} }
                                             \label{ampl}
\en
where $N_{0}$ is a suitable normalization constant dictated by
unitarity and given explicitly by:
\[
N_{0}= \frac{2}{\pi}\,\, g_{D}^{2}
\]
with $g_{D}$ being the $D$-dimensional coupling constant \cite{W}, \cite{BCF},
\cite{ACV},
related to the $26$-dimensional one and to the compactification radius by:
\[
g_{D}^{2} = g^{2}_{26} (2 \pi R)^{-26+D} .
\]
Furthermore:
\[
n_{ij} = \frac{\a'}{2} k_{R,i} \cdot k_{R,j}
\]
\[
\bar{n}_{ij} = \frac{\a'}{2} k_{L,i} \cdot k_{L,j}  .
\]
Eq. (\ref{ampl}) has been obtained by performing an average on the
polarizations and a sum on all the possible values of the roots of the
Lie algebra is understood; both of these operations make the sum of
all the terms envolving $\xi$ and
$\bar{\xi}$ null.

Eq. (\ref{ampl}) can be considered as a generalization of the dual
Virasoro amplitude; its
dependence on the lattice
is entirely contained in the variables $n_{ij}$. These are characterized
by taking integer values, as we have previously seen as a consequence of
our choice of the lattice.
On the other hand, the conservation of
the winding numbers and the compactified momenta \cite{CS} imply
the following constraints:
\[
\hat{k}_{R,1}+\hat{k}_{R,2}+\hat{k}_{R,3}+\hat{k}_{R,4}=0
\]
from which
\eq
n_{13}+n_{23}+n_{34} = - 2  \label{constr1}
\en
and
\[
n_{ij}=\frac{\a'}{2} \hat{k}_{R,i} \cdot \hat{k}_{R,j}
 \leq \frac{\a'}{2} |\hat{k}_{R,i}|^{2} = 2
\]
i.e.
\eq
-2 \leq n_{ij} \leq 2  \label{constr2}
\en
Analogously $-2 \leq \bar{n}_{ij} \leq 2$. Furthermore, since the
lattice can be chosen in such a way that the vectors $\hat{k}_{R,i}$
are roots of the Lie algebra used in constructing the lattice,
the scalar product between two of them must be integer.

By using standard techniques it is possible to write the eq. (\ref{ampl})
in the following form:
\eq
A_{tree}(s,t,u) = N_{0} \pi
\frac{\Gamma \left[ - \frac{\a'}{4}s + n_{23} + 1 \right]}
{\Gamma \left[ \frac{\a'}{4}s - \bar{n}_{23} \right]}
\frac{\Gamma \left[ - \frac{\a'}{4}t + n_{34} + 1 \right]}
{\Gamma \left[ \frac{\a'}{4}t - \bar{n}_{34} \right]}
 \frac{\Gamma \left[ - \frac{\a'}{4}u + \bar{n}_{13} + 1 \right]}
{\Gamma \left[ \frac{\a'}{4}u - n_{13}\right]} .  \label{A}
\en
Eq. (\ref{A}) holds if at least two of the differences $n_{ij} - \bar{n}_{ij}$
are non negative. Otherwise, one gets:
\eq
A_{tree}(s,t,u)= N_{0} \pi
\frac{ \Gamma \left[ -\frac{\a'}{4}s+\bar{n}_{23}+1 \right]}{\Gamma
\left[ \frac{\a'}{4}s-n_{23} \right]}
\frac{ \Gamma \left[ -\frac{\a'}{4}t + \bar{n}_{34} + 1
\right]}{\Gamma \left[ \frac{\a'}{4}t-n_{34} \right]}
\frac{ \Gamma \left[ -\frac{\a'}{4}u+n_{13}+1 \right]}{\Gamma
\left[ \frac{\a'}{4}u-\bar{n}_{13} \right]} .
                                          \label{AA}
\en
The amplitude (\ref{AA}) corresponds to (\ref{A}) in which the variables
$n_{ij}$ and $\bar{n}_{ij}$ are interchanged.

The amplitude (\ref{A}) has poles compatible with the mass formula
(\ref{const1}).
Analogously the amplitude (\ref{AA}) has poles consistent with the
constraint (\ref{const2}).

We are now interested in performing the limit $\a'
\rightarrow 0$ of the
scattering amplitudes (\ref{A}) [or, equivalently, of (\ref{AA})].
Taking into account
the possible values of the variables $n_{ij}$ and by using the analytic
properties of
the $\Gamma$-function, it is straightforward to obtain the following
result in the limit $\a' \rightarrow 0$:
\[
A_{tree}(s,t,u) = A_{tree}^{(\lambda \phi^{4})} + A_{tree}^{(\lambda \phi^{3})}
+ A_{tree}^{(spin\,\,\,1)} + A_{tree}^{(spin\,\,\,2)}
\]
with
\eq
A_{tree}^{(\lambda \phi^{4})} \sim g^{2}_{D}              \label{const}
\en
\eq
A_{tree}^{(\lambda \phi^{3})} =
- 2 \,\,\, \frac{4}{\a'} \,\,\, g_{D}^{2}
\left[ \frac{1}{s} + \frac{1}{t} + \frac{1}{u} \right]
\label{sc}
\en
\eq
A_{tree}^{(spin\,\,\,1)} = 2 \,\,\, g_{D}^{2} \,\,\,
\left[ \frac{u+t}{s} + \frac{s+u}{t} + \frac{s+t}{u} \right]
\label{sp1}
\en
\eq
A_{tree}^{(spin\,\,\,2)} =
2 \,\,\, \frac{\a'}{4} \,\,\, g_{D}^{2} \,\,\,
\left[ \frac{tu}{s} + \frac{su}{t} + \frac{st}{u} \right]
\label{sp2}
\en

The amplitudes (\ref{sc}), (\ref{sp1}) and (\ref{sp2}) represent
scalars
interacting with the exchange respectively of a scalar ($\lambda \phi^{3}$
theory, with $\lambda^{2} \equiv 2 \frac{4}{\a'} g^{2}_{D} $) ,
spin 1 (``scalar electrodynamics'') and spin 2 particle (quantum gravity).
The constant value (\ref{const}) can be interpreted as
a tree diagram of an ``effective'' $\lambda \phi^{4}$ theory, coming from a
tree
diagram of a $\lambda \phi^{3}$ theory on which the double limit produces
an ultralocal limit of the propagator.

{}From the comparison of the amplitude (\ref{sp2}) with the analogous one
computed in quantum gravity, we can obtain a relationship between the
string coupling
constant $g_{D}$ and the gravitational coupling constant $G_{N}$ in
$D=4$ dimensions:
\[
g^{2}_{4} = \frac{ 16 \pi G_{N}}{\a'}.
\]
This expression coincides with the one already known in literature
\cite{ACV}.

In conclusion we have explicitly shown that the compactification of the
closed bosonic string theory reproduces only in the double limit $\a'
\rightarrow 0$ and $R \rightarrow 0$, at the tree level,
the ordinary field theories. We would like to stress here that in
our work specifying the lattice has resulted to be unnecessary:
hence, at least at that level, compactification does not influence the
low-energy limit.

We acknowledge G. Cristofano, P. Di Vecchia, G. Maiella,
R. Musto and R. Pettorino for discussions and a critical reading of
the manuscript. Nordita is acknowledged for their kind hospitality
in some stages of this work. Finally, one of us (R. M.) particularly
thank V. Marotta for fruitful discussions on lattices and Lie algebras.

\section*{Figure Captions.}
\begin{description}

\item{FIG. 1} Scattering of four scalar particles.

\end{description}

\newpage

\begin{center}

\begin{tabular}{|c|c||c|c|} \hline
{}~~~~ & ~~~~~~~~ & ~~~~~~~~~ & ~~~~~~~~~~~~~~\\
$N$ & $\bar{N}$ & $p^{2}_{R}$ & $p^{2}_{L}$ \\
{}~~~~ & ~~~~~~~  & ~~~~~~~~  & ~~~~~~~~~~~~~~\\
\hline
{}~~~~&~~~~&~~~~~~~~~&~~~~~~~~~\\
0   &  0 & $4/\a'$ & $4/\a'$ \\
{}~~~~&~~~~&~~~~~~~~~&~~~~~~~~~\\
0   & 1 &  $4/\a'$ &   0     \\
{}~~~~&~~~~&~~~~~~~~&~~~~~~~~~~\\
1   & 0 &   0 &   $ 4/\a'$    \\
{}~~~~&~~~&~~~~~~~~&~~~~~~~~~~~\\
1   & 1 &  0 & 0 \\
{}~~~~&~~~&~~~~~~~~&~~~~~~~~~~~\\ \hline
\end{tabular}

\vspace{0.5cm}
{\bf Tab. 1}

\end{center}


\newpage

\begin{thebibliography}{99}
\bibitem{GSW} M.B. Green, J.H. Schwarz and E. Witten, {\em Superstring
Theory}, vol. I, Cambridge University Press (1987).
\bibitem{LT} D. L\"{u}st and S. Theisen, {\em Lectures on String Theory},
Lecture Notes in Physics,
Springer-Verlag (1989).
\bibitem{LSW} W. Lerche, A.N. Schellekens and N.P. Warner, Phys. Rep.
{\bf 177} (1989) 1.
\bibitem{Y} T. Yoneya, {\em Lett. Nuovo Cim.} {\bf 8} (1973) 951.
\bibitem{GS} M.B. Green, J.H. Schwarz and L. Brink {\em Nucl. Phys.} {\bf B198}
(1982) 474.
\bibitem{PDV} P. Di Vecchia, F. Pezzella, M. Frau, K. Hornfeck, A. Lerda
and S. Sciuto, {\em Nucl. Phys.} {\bf B322} (1989) 317; \\
A. Clarizia and F. Pezzella, {\em Nucl. Phys.} {\bf B301} (1988) 499.
\bibitem{CS} E. Cremmer and J. Scherk, {\em Nucl. Phys.} {\bf B103} (1976) 399.
\bibitem{W} S. Weinberg, {\em Phys. Lett.} {\bf 156B} (1985) 309.
\bibitem{BCF} A. Bellini, G. Cristofano, M. Fabbrichesi and K. Roland, {\em
Nucl. Phys.} {\bf B356} (1991) 69.
\bibitem{ACV} D. Amati, M. Ciafaloni and G. Veneziano, {\em Phys. Lett.} {\bf
197B} (1987)
129; Int. J. Mod. Phys. {\bf A3} (1988) 1615.
\end{thebibliography}
\end{document}